\documentclass{mn2e}
\newcommand{\hide}[1]{} 
\usepackage{times,graphicx,amssymb,color,pifont}
\def\arcsec{\hbox{$^{\prime\prime}$}}
\newcommand{\xmark}{\ding{55}}%
\newcommand{\cmark}{\ding{51}}%
\title[Orbits of the terrestrial planets]{Constraining the primordial orbits of the Terrestrial Planets}
\author[R. Brasser et al.]{R. Brasser$^1$, K. J. Walsh$^2$ and D. Nesvorn\'{y}$^2$\\
$^1$ Institute for Astronomy and Astrophysics, Academia Sinica, Taipei 10617, Taiwan\\ 
$^2$ Southwest Research Institute, 1050 Walnut St., Suite 300, Boulder, CO, 80302, USA}
\begin{document}
\maketitle
\begin{abstract}
Evidence in the Solar System suggests that the giant planets underwent an epoch of radial migration that was very rapid, with
an e-folding timescale shorter than 1~Myr. It is probable that the cause of this migration was that the giant planets experienced an
orbital instability that caused them to encounter each other, resulting in radial migration. A promising and heavily studied way to
accomplish such a fast migration is for Jupiter to have scattered one of the ice giants outwards; this event has been called the
`jumping Jupiter' scenario. Several works suggest that this dynamical instability occurred `late', long after all the planets had
formed and the solar nebula had dissipated. Assuming that the terrestrial planets had already formed, then their orbits would have
been affected by the migration of the giant planets as many powerful resonances would sweep through the terrestrial planet region.
This raises two questions. First, what is the expected increase in dynamical excitement of the terrestrial planet orbits caused by
{ late} and very fast giant planet migration? And second, { assuming the migration occurred late}, can we use this
migration of the giant planets to obtain information on the primordial orbits of the terrestrial planets? In this work we attempt to
answer both of these questions using numerical simulations. We directly model a large number of terrestrial planet systems and their
response to the smooth migration of Jupiter and Saturn, and also two jumping Jupiter simulations. We study the total dynamical
excitement of the terrestrial planet system with the Angular Momentum Deficit (AMD) value, including the way it is shared among the
planets. We conclude that to reproduce the current AMD with a reasonable probability ($\sim$20\%) after { late} rapid giant planet
migration { and a favourable jumping Jupiter evolution}, the primordial AMD should have been lower than $\sim$70\% of the current
value, { but higher than 10\%}. We find that a { late} giant planet migration scenario that initially had five giant planets
rather than four had a higher probability to satisfy the orbital constraints of the terrestrial planets. { Assuming late migration} we
predict that Mars was initially on an eccentric and inclined orbit while the orbits of Mercury, Venus and Earth were more circular and
coplanar. The lower primordial dynamical excitement and the peculiar partitioning between planets impose new constraints for
terrestrial planet formation simulations. 
\end{abstract} 
\begin{keywords}
Solar System: general
\end{keywords}

\section{Introduction}
It is thought that the giant planets did not form where they are today
but instead have migrated in the past (e.g. Fernandez \& Ip, 1984;
Hahn \& Malhotra, 1999). It is also thought that this migration was
not smooth but that instead the outer planets suffered a dynamical
instability where at least one ice giant was scattered by Jupiter and
Saturn (Thommes et al., 1999). This dynamical instability of the giant
planets, and subsequent mutual scattering among them, is the most
likely way to explain their current eccentricities and inclinations
(Tsiganis et al., 2005; Morbidelli et al., 2009). This episode of
mutual scattering ensures that the migration of Jupiter and Saturn was
fast enough to keep the asteroid belt stable (Minton \& Malhotra, 2009; Morbidelli et al.,
2010), and, { if this migration occurred late, also} the terrestrial planets 
(Brasser et al., 2009). A dynamical instability in the outer solar system could also
explain many additional features of the outer solar system we observe
today: the capture and orbital properties of Jupiter's Trojans
(Morbidelli et al., 2005; { Nesvorn\'{y} et al., 2013}), the orbital properties and structure of
the Kuiper Belt (Levison et al., 2008), the capture of the
irregular satellites of the giant planets (Nesvorn\'{y} et al.,
2007) and, { if the migration occurred late}, the Late Heavy Bombardment of the terrestrial planets 
(Gomes et al., 2005; Bottke et al., 2012), Here we { assume throughout that the migration of the giant planets coincided with the
Late Heavy Bombardment and thus occurred after the formation of the terrestrial planets. We} study the consequence of { this late}
dynamical instability of the giant planets on the inner solar system. \\

Brasser et al. (2009) attempted to reproduce the current secular
architecture of the terrestrial planets in response to { late} giant 
planet migration. They showed that the migration of Jupiter and
Saturn needed to have been very fast, otherwise the eccentricities of
the terrestrial planets would have been excited to values much higher
than they are today. The eccentricity excitation is caused by a
powerful secular resonance between the terrestrial planets and
Jupiter. As Jupiter and Saturn drift apart Jupiter's proper precession
frequency, $g_5$, decreases and crosses the proper frequencies
associated with the terrestrial planets (Brasser et al., 2009; Agnor
\& Lin, 2012). Thus, the sweeping of $g_5$ causes the terrestrial
planets to experience the secular resonances $g_5=g_n$, with
$n=1\ldots4$. Here $g_n$ are the proper eccentricity eigenfrequencies
of the terrestrial planets. Both works demonstrated that the crossing
of the resonances $g_5=g_2$ and $g_5=g_1$ caused the greatest
eccentricity increase in Mercury, Venus and Earth because they were
crossed slowly. The resonances with $g_1$ and $g_2$ occur when the
Saturn to Jupiter period ratio is $P_S/P_J \sim 2.25$ and $P_S/P_J
\sim 2.15$. \\


In Brasser et al. (2009) two solutions were presented for solving the
problem of keeping the excitation of the orbits of the terrestrial
planets at their current level. The first mechanism is one in which
the secular resonances $g_5=g_2$ and $g_5=g_1$ had a phasing that
resulted in the eccentricity either decreasing or that the
eccentricity increase in each planet was very small. The probability
of this phasing was estimated at approximately 10\% (see also Agnor \&
Lin, 2012). The second mechanism was for the migration of Jupiter and
Saturn to have proceeded very quickly by Jupiter scattering one of the
ice giants outwards. Energy and angular momentum conservation causes
Jupiter to move inwards as it scatters the ice giant outwards and
increases the period ratio with Saturn. The typical time scale of
Jupiter's migration is 100~kyr. The scattering scenario for Jupiter's
migration was dubbed the 'jumping Jupiter' scenario because the
semi-major axis of Jupiter appears to undergo a sudden decrease, 
increasing the period ratio with Saturn. Minton \& Malhotra (2009)
advocated a short migration time scale based on the structure of the
asteroid belt. This led Morbidelli et al. (2010) to expose a 
model asteroid belt to a jumping Jupiter simulation. They
concluded that the migration of the giant planets had to be of this
type because it is the only known physical mechanism that can drive
such fast migration. This result was further supported by Walsh \&
Morbidelli (2011), who demonstrated that the migration of Jupiter had
to be fast whether this occurred { late} i.e. at the time of the Late Heavy
Bombardment, or right after the gas disc had dissipated. { Agnor \& Lin (2012) were aware of the difficulty of keeping the
terrestrial system stable as the gas giants migrated and for this reason they advocated an early migration, occurring before the
terrestrial system had fully formed.}\\

In a separate study Agnor \& Lin (2012) investigated the effect of the
migration of the giant planets on the terrestrial planets. They kept
track of changes in the terrestrial planet's Angular Momentum
Deficit (AMD), which is a measure of a system's deviation from being
perfectly circular and coplanar (Laskar, 1997). { Here we adopted a dimensionless variation, though we shall continue to refer to
it as `AMD' for simplicity. It is defined as (e.g. Raymond et al., 2009)}

\begin{equation}
 {\rm{AMD}} = \frac{\sum_n m_n\sqrt{a}_n(1-\sqrt{1-e_n^2}\cos i_n)}{\sum_n m_n \sqrt{a}_n},
\end{equation}
where $m$ is the mass of plane $n$ in units of the Solar mass, $a$ is the semi-major axis of said planet, $e$ is its eccentricity and
$i$ is its inclination with respect to a reference plane (in our case, the invariable plane). Agnor \& Lin (2012)
noticed that the current distribution of the eccentricity { contribution to} the AMD is mostly contained in the components
corresponding to Mercury and Mars. These two components have a combined total value of
85\% of the system's AMD. This led Angnor \& Lin (2012) to
conclude that { late} migration of Jupiter and Saturn had to occur with an
e-folding time scale $\tau \ll$ 1~Myr, otherwise the AMD of the
terrestrial planets would be incompatible with its current
value. This result is in agreement with Brasser et al. (2009). From
their numerical experiments Agnor \& Lin (2012) find that the
excitation of the eccentricities of the terrestrial planets scales as
$\tau^{1/2}$. However, they conclude that when $\tau \lesssim 0.1$~Myr
the excitation imposed on the terrestrial planets is independent of
$\tau$ because the excitation is impulsive rather than adiabatic. In
other words, for values of $\tau$ shorter than 0.1~Myr the AMD increase
of the terrestrial planets is independent of $\tau$ and is equal to a
constant value. Here we try to determine the magnitude of the
excitation of the AMD of the terrestrial planets during this fast
migration.\\

Agnor \& Lin (2012) also investigated the most likely primordial
orbits of the terrestrial planets that are consistent with this fast { late}
migration. They concluded that the primordial amplitudes of the
eccentricity modes associated with Venus and Earth had to be nearly 0.
On the other hand the primordial amplitudes of the eccentricity modes
corresponding to Mercury and Mars were comparable to their current
values. These results suggest that Mercury and Mars were already
eccentric (and possibly inclined) before giant planet migration while
Earth and Venus obtained their eccentricities after the migration of
the gas giants. \\

In summary, multiple studies point towards a very rapid migration of
the giant planets, most likely of the jumping Jupiter variety. 
{ If this migration occurred late} this raises two questions. 
First, the terrestrial planets are excited even
if the period ratio jumps far enough (beyond 2.3) and the resonances
$g_5=g_1$ and $g_5=g_2$ are not activated. However, it is not {\it a
  priori} clear how much the terrestrial planets are excited if the
period ratio jumps beyond 2.3. This raises the issue of what were the
initial orbits of terrestrial planets that could meet these
constraints. Second, does a jump that does not destabilise the
terrestrial planets occur in a statistically significant number of
cases? The Jupiter-Saturn period ratio needs to `jump' from $\sim$1.5
to beyond 2.3 and then also avoid moving past the current value of
2.49.\\

In response to the first question, in this paper we determine how late
giant planet migration changed the orbits of the terrestrial planets
with the use of numerical simulations. The aim is to provide an upper
limit on the primordial AMD of the terrestrial planets. Knowing the
possible range of terrestrial planet orbits before the { late} migration of
the giant planets { could impose} a constraint for models of
terrestrial planet accretion. { While current terrestrial planet formation simulations are capable of generating systems whose AMD
and mass distribution matches the current terrestrials (e.g. Hansen, 2009; Walsh et al., 2011; Raymond et al., 2009), it is possible
that late giant planet migration substantially increased the AMD of the terrestrials. Current simulations are unable to form a cold
terrestrial system with the right mass distribution. Here we aim to quantify this AMD increase and thus impose a possible new target
that terrestrial planet formation simulations should reproduce.} We set up mock terrestrial planet
systems with different initial AMDs and phasing, and expose them to the
instability of the giant planets. We then statistically analyse the
results and determine the resulting orbital structure and AMD
values.\\

Regarding the second question Brasser et al. (2009) concluded that the
probability of a jumping Jupiter simulation that kept all four giant
planets, and had the period ratio rapidly increase to 2.3 or higher,
was very low. This led Nesvorn\'{y} (2011) to suggest the solar system
may have contained a third ice giant that was ejected during the late
dynamical instability (see also Batygin et al.  2012). This idea
was followed up by Nesvorn\'{y} \& Morbidelli (2012). They ran 10\,000
simulations of late giant planet instabilities from a large variety of
initial conditions and performed 30 to 100 simulations per set of
initial conditions to account for stochastic effects. They included
simulations with four, five and six giant planets and imposed four
stringent constraints each simulation should fulfil to be considered
successful. One of these constraints was that the Jupiter-Saturn
period ratio should jump to 2.3 or higher but end below 2.5. The large
number of simulations for each set of initial conditions allowed them
to quantify the probability of the outcome adhering to all four
constraints. With initially four planets the probability was found to
be less than 1\%, while it increased to 5\% for certain configurations
of five planets. Thus, it appears that the 5-planet model is a more
promising avenue to reproduce the current configuration of the outer
planets than a 4-planet case.\\

The work by Nesvorn\'{y} \& Morbidelli (2012) was just the first
simple attempt to trace the dynamical history of the outer planets
because the terrestrial planets were not included in their
simulations. Instead, they used the simple period ratio constraints
from Brasser et al. (2009) and assumed that the terrestrial planets
would survive the migration if these were satisfied. Here we include
the terrestrial planets in the simulations and expose them to a few cases from the
above works to check their behaviour more explicitly.\\

The aim and methodology of this paper differ from those of Brasser et
al. (2009) and Agnor \& Lin (2012) in several ways. First, Brasser et
al. (2009) only investigated whether the eccentricities of the
terrestrial planets could be kept below or at their current values if
the gas giants migrated quickly. They did not investigate the range of
possible outcomes when the initial phasing of the terrestrials
differed at the time of the instability. Agnor \& Lin (2012) went
further than Brasser et al. (2009) and performed simulations
to obtain a more stringent limit on the time scale of the migration of
Jupiter and Saturn. However, their simulations were all of the smooth
migration variety and were controlled by having a pre-determined
finite migration range and time scale. While they measured the
probability of keeping the eccentricities of the terrestrial planets
below a certain threshold value as a function of the migration
e-folding time, they only did so for each planet individually rather
than focus on the terrestrials as a system. The approach taken here is
to consider the excitation of the whole terrestrial system by
measuring the difference between the AMD after and before migration,
and the mean value and variance of the final AMD. From there we
determine the most likely primordial value of the AMD. We find that if
the gas giants' period ratio jumps from 1.5 to beyond 2.3 and remains 
below 2.5, then the median final AMD of the terrestrial planets
equals the current value if they were initially dynamically cold. We
also find that Mars had to be more excited than the other three.\\

This paper is divided as follows. In Section 2 we present the details
of our numerical simulations and their initial conditions. This is
followed by the results in Section 3. Section 4 is reserved for the
discussion and implications of this work and our conclusions are drawn
in the last section.

\section{Methods}
The current study is based on a large set of numerical simulations
where we subject the terrestrial planets to the effects of the
evolution of the migrating giant planets. In principle, we prefer to
subject the terrestrial planets to a series of Nice model simulations
with fast migration of the giant planets. However, the evolution of
the giant planets is chaotic and this makes it difficult to quantify
changes in the orbits of the terrestrial planets.\\

Thus we first performed a series of simulations where we subject the
terrestrial planets to smooth migration of the gas giants with various
initial values of their period ratio. We set the e-folding time for
their smooth migration at $\tau = 3$~Myr and at 1~Myr, where 1~Myr is 
the shortest time scale we have witnessed to occur in
self-consistent smooth migration simulations of the giant planets due
to planetesimal scattering; the typical time scale is 3 to 5~Myr (Hahn
\& Malhotra, 1999; Morbidelli et al., 2010). We also made sure that
the final amplitude of the $g_5$ mode in Jupiter, $e_{55}$, is close
(0.041) to its present value (0.043).\\

\begin{figure}
\resizebox{\hsize}{!}{\includegraphics[angle=-90]{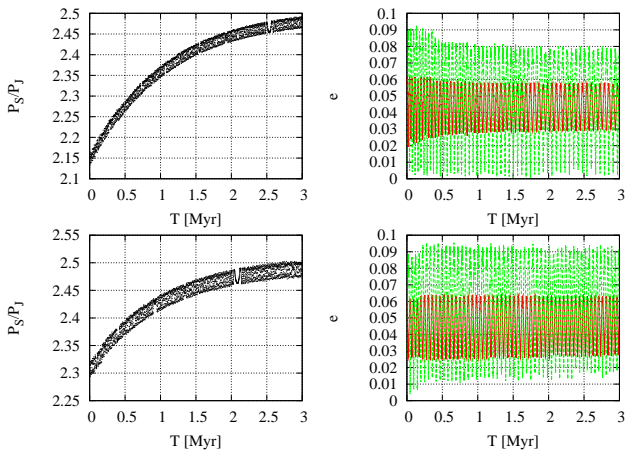}}
\caption{The period ratio $P_S/P_J$ (left panels) and eccentricities
  of Jupiter (red) and Saturn (green) (right panels) for two of the
  smooth migration runs. The other simulations have a similar
  evolution.}
\label{pspj}
\end{figure}

Brasser et al. (2009) and Agnor \& Lin (2012) show that during the
migration the terrestrial planets experience the effects of the
resonances $g_5=g_2$ at $P_S/P_J \sim 2.15$ and $g_5=g_1$ at $P_S/P_J
\sim 2.25$. Agnor \& Lin (2012) further conclude that the current
amplitudes of the eccentricity eigenmodes of Mercury and Venus,
$e_{11}$ and $e_{22}$, can only be reproduced when these resonances
are crossed on time scales $\tau_2 \lesssim 0.05$~Myr and $\tau_1
\lesssim 0.7$~Myr, respectively. Therefore we ran a large number of
smooth migration experiments where we place Jupiter and Saturn on
orbits with $P_S/P_J$ ranging from 2.15 to 2.3 in increments of
0.01. The initial conditions mimic a jump to this period
ratio. Examples of the evolution of $P_S/P_J$ and the eccentricities
of Jupiter and Saturn during these smooth migration simulations are
given in Fig.~\ref{pspj}.\\

We quantify the final eccentricities of Mercury and Venus because
these two planets are the most vulnerable to the sweeping of $g_5$.
We consider a simulation outcome to be successful if the maximum
eccentricity of Mercury remained below 0.35 and that of Venus below
0.09. These upper limits on the eccentricities of Mercury and Venus
are based on the results of Laskar (2008), who showed that long-term
chaotic diffusion of the eccentricities and inclinations of the
terrestrial planets can significantly alter their mean values from the
current ones. Over the age of the solar system Mercury has a 50\%
chance to have its eccentricity increased above 0.35 when starting at
the current value, while Venus has a 50\% probability of its
eccentricity exceeding 0.09. A second constraint on Mercury's
eccentricity comes from its rotation: if its eccentricity had exceeded
0.325 for a long time it would most likely have been trapped in the
2:1 spin-orbit resonance rather than in the current 3:2 (Correia \&
Laskar, 2010).\\

The smooth migration simulations were supplemented with Nice-model
jumping Jupiter simulations. In this study we show the results of one
4-planet Nice model simulation (`classical Nice') and one 5-planet
case (Nesvorn\'{y} \& Morbidelli, 2012). The numerical simulations
were performed with SWIFT RMVS3 (Levison \& Duncan, 1994), which was
modified to read in the evolution of the giant planets and compute
their intermediate positions by interpolation (Petit et al., 2001;
Brasser et al., 2009; Morbidelli et al., 2010). All the planets from
Mercury to Neptune were included in the Nice model simulations, while
we included only Mercury to Saturn in the smooth migration runs. We
added the effects of general relativity { in all our simulations} according to the method
described in Nobili \& Roxburgh (1986). { This consisted of adding the effects of a disturbing potential to SWIFT whose effect
generates the correct perihelion precession, but does not reproduce the increased orbital frequency (Saha \& Tremaine, 1994). This
potential is
\begin{equation}
V_{\rm GR}=-3\Bigl(\frac{GM_{\odot}}{c}\Bigr)^2\frac{a}{r^3},
\end{equation}
where $G$ is the gravitational constant, $M_{\odot}$ is the Solar mass, $c$ is the speed of light and $r$ is the planet-Sun distance.}
In all of our simulations the time step was set at 0.02~yr (approximately 7~days). The series of
simulations consisted of the following.\\

First we determined the new initial orbits of the terrestrial planets { using the AMD as the independent variable. We chose to base
our system on its AMD value and the share in each planet rather than individual orbits because this greatly simplifies the subsequent
analysis.} We took the current orbits of the terrestrial planets { with respect to the Solar System's invariable plane from the
IMCCE's ephermerides website\footnote{http://www.imcce.fr}} and computed the instantaneous AMD and the share in each planet. The share
of the AMD in each planet is currently 34\% in Mercury, 20\% in Venus, 21\% in Earth and 25\% in Mars, { and is computed as}
\begin{equation}
 f_n=\frac{m_n \sqrt{a}_n(1-\sqrt{1-e_n^2}\cos i_n)}{\sum_k m_k \sqrt{a_k} (1-\sqrt{1-e_k^2}\cos i_k)}.
 \label{fracp}
\end{equation}
The current AMD value and share in each planet forms
our base orbit set. { We verified with numerical simulations that the current AMD and the share in each planet are representative of
their long-term average values.} Second, for each simulation the initial value of the AMD was scaled from the current one. This simple
approach allowed us to mimic systems whose primordial AMD was lower or higher
than the current value but with the same partitioning among the
planets. For the Nice model simulations we scaled the AMD ranging from
0.1 to 2.2 times the current value in increments of 0.3. The higher
values were chosen to determine if destructive interference during
giant planet migration could lower the primordial AMD. For the smooth
migration experiments this scaling was 0.1, 0.5 and 1.0.\\

Third, the initial eccentricities and inclinations of the terrestrial
planets were calculated by assuming that $e=\sin i$ and by keeping the
semi-major axes fixed at their current values. For the smooth
migration experiments we also ran a separate set where all the AMD was
in Mercury's eccentricity (so that its initial values were 0.088,
0.195 and 0.276 respectively). The longitude of the ascending node
($\Omega$), argument of pericentre ($\omega$) and mean anomaly ($M$)
of each terrestrial planet were chosen at random on the interval 0 to
360$^\circ$. This randomisation was done to account for the planets
having different phasing when the gas giants migrate. \\

Fourth, the newly-generated terrestrial systems were subjected to a
jumping Jupiter or smooth migration evolution. For the jumping Jupiter
cases the final system was integrated with SWIFT RMVS3 for an
additional 2~Myr to obtain the averaged final AMD. Last, for the Nice
model cases the averaged final AMD value was recorded, together with
the {\it averaged} share of the AMD that each planet possesses. { These 
averages were obtained over the last quarter of the simulations.} We
also calculated $\langle e \rangle/\langle \sin i \rangle$ for each
planet. For the smooth migration experiments we computed the
cumulative distributions of Mercury's and Venus' minimum, mean and
maximum eccentricity and the probability that their maxima are below
0.35 and 0.09 respectively. We ran 300 simulations for each value of
the initial AMD to account for statistical effects and phasing. The
evolution of the gas giants was kept the same for each simulation. The
integrator SWIFT RMVS3 cannot handle close encounters between the
planets (Levison \& Duncan, 1994) and thus a simulation was stopped
when a pair of planets encountered each other's Hill spheres or when
they were farther than 500~AU from the Sun. All simulations were
performed on the ASIAA HTCondor pool.\\

\section{Results}
In this section we present the results of our numerical
simulations. Rather than discuss all the possible outcomes from each
set of simulations, we shall take a more statistical approach. \\

\subsection{Smooth migration experiments}
We first studied the effect of smooth migration of Jupiter and
Saturn on the terrestrial planets. We focus our attention on Mercury
and Venus because these two planets are the most vulnerable to the
secular sweeping of $g_5$ through the terrestrial region. The goal of
these simulations is to establish the final eccentricities of Mercury
and Venus as a function of initial $P_S/P_J$ and AMD, and the
probability that the maximum eccentricity of Mercury remains below
0.35 and that of Venus below 0.09. 

\subsubsection{Typical migration: $\tau=$3~Myr}
We first simulate Jupiter and Saturn's migration with an e-folding time scale of 
$\tau=3$~Myr, which is a typical value found in smooth
migration simulations (Hahn \& Malhotra, 1999). We kept the amplitude
of Jupiter's eccentricity eigenmode term, $e_{55}$, as close as
possible to its current value to best mimic the effect of the sweeping
of the $g_5$ frequency. We placed Jupiter and Saturn on orbits with
initial period ratio between 2.15 and 2.3 in increments of 0.01. The
AMD of the terrestrial planets was then scaled to either a tenth
(0.1), half (0.5) or equal to the current value. The AMD was either shared 
among the planets as is found today, or put entirely into the orbit of Mercury. 
We have presented the results in a series of four figures.\\

\begin{figure}
\resizebox{\hsize}{!}{\includegraphics[angle=-90]{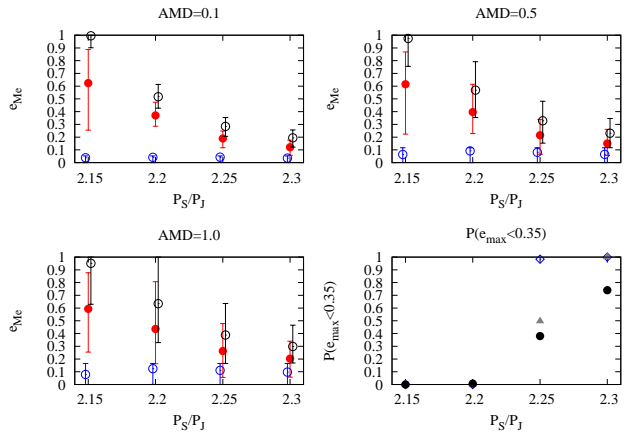}}
\caption{Average minimum eccentricity (blue circles), average mean
  eccentricity (red bullets) and average maximum eccentricity (black
  circles) of Mercury as a function of initial Jupiter-Saturn period
  ratio (top panels and bottom-left panel). The error bars depict the
  range of eccentricities. Three different initial terrestrial
    planet AMD values are shown (as a fraction of the current value);
    0.1 (upper left), 0.5 (upper right) and 1.0 (lower left). The
  bottom-right panel depicts the probability for Mercury's average maximum
  eccentricity to remain below 0.35 as a function of initial period
  ratio. The blue { diamonds} are for low AMD, grey triangles correspond
  to the run where the AMD was half of the current value case and the
  bullets are for cases where the AMD equals the current one. { For clarity we only depict results in 0.05 increments of initial
period ratio.}}
\label{em3m}
\end{figure}

\begin{figure}
\resizebox{\hsize}{!}{\includegraphics[angle=-90]{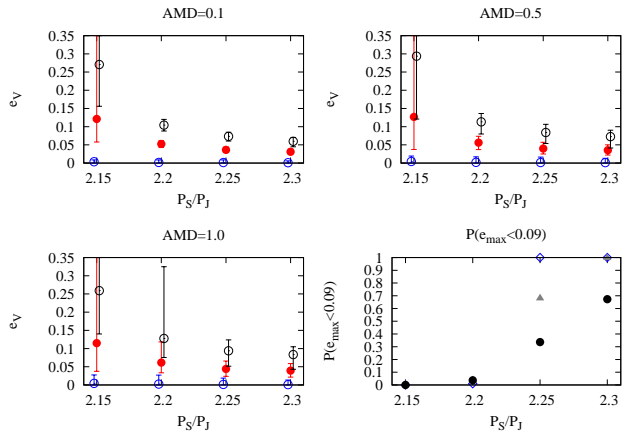}}
\caption{Same as Fig.~\ref{em3m} but here the eccentricity of Venus is plotted. { For clarity we only depict results in 0.05
increments of initial period ratio.}}
\label{ev3m}
\end{figure}

Figure~\ref{em3m} plots the {\it averages} of the minimum eccentricity
(blue), the mean eccentricity (red) and the maximum eccentricity
(black) of Mercury as a function of initial $P_S/P_J$ for several
initial terrestrial planet AMD. The initial AMD sharing among
each planet was kept at the current one. The error bars depict
the variation of each quantity, and the averages were computed
from data taken during the last 0.5~Myr of the migration
simulations. The data points for the average minimum and maximum
eccentricity have a slight horizontal offset from the mean for
clarity. There are two visible trends.\\

First, the final eccentricity of Mercury decreased as the initial
$P_S/P_J$ was increased from 2.15 to 2.3. When the initial
period ratio is at 2.15, both the $g_5=g_2$ and $g_5=g_1$ resonances
are crossed, but when the initial period ratio is beyond 2.2,
only $g_5=g_1$ is crossed. The second trend is that the mean values
and the spread increase with initial AMD (comparing the upper
  left with the upper right and the lower left panels of the two
  figures). The spread in the average mean and average maximum values
decrease slightly with $P_S/P_J$.\\

The bottom-right panel of Fig.~\ref{em3m} depicts the probability that
the maximum eccentricity of Mercury stays below 0.35. The blue squares
correspond to the case with low initial AMD (0.1), the grey
triangles are for half AMD (0.5) and the bullets correspond to
the cases with current AMD (1.0). The probability increases with
increasing period ratio, and reaches near unity for the low-AMD case
(0.1) when the initial period ratio is higher than 2.25. { Larger values of initial AMD 
require a period ratio of 2.3 or above.} When the initial period ratio is 
lower than 2.2 the chance of keeping Mercury's eccentricity in check is zero.\\

The same procedure is repeated for Venus in Fig.~\ref{ev3m}. We
restricted the calculation to the probability of Venus' maximum
eccentricity staying below 0.09 rather than repeating Agnor \& Lin
(2012) and requiring that the amplitude of Venus' eccentricity
eigenmode $e_{22}<0.02$. The probability of acceptable eccentricity for Venus as 
a function of $P_S/P_J$ is very similar to that found for Mercury (a
comparison of the lower right panels of Figures \ref{em3m} and \ref{ev3m}). Probabilities for all cases were $\sim$0\% for $P_S/P_J
<$2.2, and $\gtrsim$80\% for $P_S/P_J >$2.3 for both planets; 
{ extrapolation suggests that all probabilities reach unity when the period ratio exceeds 2.35.} \\

In summary, we have demonstrated that the smooth migration of
Jupiter and Saturn with $\tau=$3~Myr is only capable of reproducing
the current eccentricities of Mercury and Venus { with a high probability} from a period ratio of
2.25 or higher if the original AMD was very low; { otherwise the period ratio should exceed 2.3.} 
However, we did not investigate the effect of changing the migration time scale. This is done below.\\

\subsubsection{Fast migration: $\tau=$1~Myr}
Above we investigated how the terrestrial planets respond to smooth
migration of Jupiter and Saturn starting from $P_S/P_J=$2.15 to 2.30
and migrating to their current period ratio. We concluded that the
eccentricities of Mercury and Venus are compatible with their current
values when the jump proceeds to 2.3 or beyond. However, we
only focused on the typical migration time scale
of $\tau=$~3~Myr. Agnor \& Lin (2012) demonstrated that the
increase in the eccentricities of the terrestrial planets scales as
$\tau^{1/2}$. However, it is not known how the spread in the
eccentricity, and thus the probability of keeping it below a specified
value, scales with $\tau$. Therefore, we have performed smooth
migration experiments where $\tau=$~1~Myr, which is the lowest value
found to occur in Nice model simulations after the period ratio
jump. This migration speed is considered to be an extreme
case. \\

\begin{figure}
\resizebox{\hsize}{!}{\includegraphics[angle=-90]{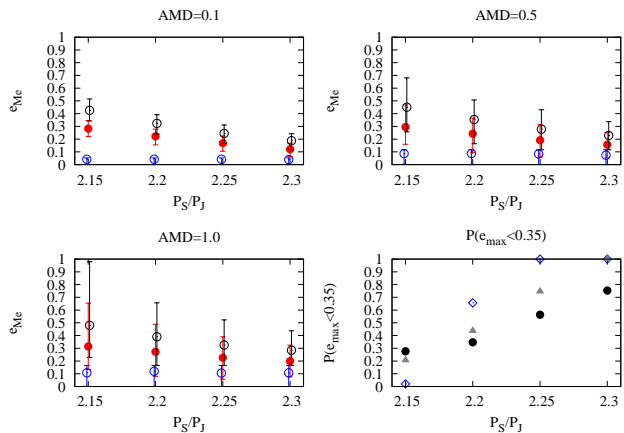}}
\caption{The same as Fig.~\ref{em3m} but now $\tau=1$~Myr.}
\label{em1m}
\end{figure}

\begin{figure}
\resizebox{\hsize}{!}{\includegraphics[angle=-90]{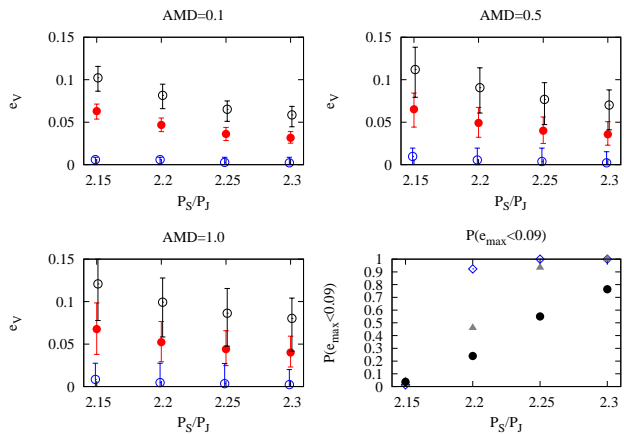}}
\caption{The same as Fig.~\ref{ev3m} but now $\tau=1$~Myr.}
\label{ev1m}
\end{figure}

The results are plotted in Figs.~\ref{em1m} and~\ref{ev1m}, and are 
similar to that found for $\tau=$3~Myr. The primary difference is 
that there is a non-zero probability for acceptable eccentricities of 
Venus and Mercury for period ratios below 2.2, whereas this was
essentially zero for the longer time scale migration. However, the spread 
in the final eccentricity does not seem to strongly depend on $\tau$.\\

In these smooth migration experiments we tested two different migration time scales. They both produce largely similar results, which
are summarised in Table \ref{tab2}. The eccentricities of both Mercury and Venus can be reproduced { with a high probability} when
i) the period ratio $P_S/P_J$ jumps directly to 2.3, ii) the period ratio $P_S/P_J$ jumps to 2.25 if the AMD was initially $\sim$0.5
of today's value or iii) the period ratio $P_S/P_J$ jumps to 2.2, the AMD was low and the migration was then very fast
($\tau \sim$~1~Myr).\\

\begin{table}
\begin{tabular}{lc|ccc | ccc|}
    \hline
  &Initial AMD  & 0.1 & 0.5 & 1.0  & 0.1 & 0.5 & 1.0 \\
$P_S/P_J$& & \multicolumn{3}{|c|}{$\tau =1$~Myr} & \multicolumn{3}{|c|}{$\tau =3$~Myr}   \\
\hline
2.15  &   & \xmark & \xmark& \xmark& \xmark& \xmark &\xmark  \\
2.20  &   &\cmark & \xmark& \xmark&  \xmark & \xmark&  \xmark \\
2.25  &   &\cmark &\cmark &\cmark   & \cmark  &\cmark & \xmark \\
2.30  &   &\cmark &\cmark & \cmark  & \cmark  &\cmark &\cmark \\
\hline
\end{tabular}
\caption{ A summary of the Figures 2-5, where a set of parameters
  received a \cmark if both Mercury and Venus had acceptable
  maximum eccentricities ($e<0.35$ and $e<0.09$ respectively) in at
  least 50\% of the simulations tested at that set of parameters. Otherwise we placed a \xmark.
\label{tab2}}
\end{table}

The above smooth migration tests were designed to mimic the evolution after
a `jump', but by starting the simulation at a given period ratio $P_S/P_J$ they
  did not explicitly model the actual jump. Therefore these results are
  likely optimistic for maintaining an acceptable system AMD for any
  given simulation. We now turn our focus on subjecting the
terrestrial system to some Nice model simulations, in order to
  quantify how the system of terrestrial planets responds to the
  actual `jump' of period ratio $P_S/P_J$.  We discuss the results
of these experiments in the next subsections.

\subsection{A 4-planet Nice model simulation}
In the previous subsection we have demonstrated that it may be
possible to reproduce the current eccentricities of Mercury and Venus
after the giant planets underwent a late instability, provided that
some criteria are met about the evolution of the gas giants. The reproduction
becomes more likely when the orbits of the terrestrials were
dynamically colder than today. Here we demonstrate a case of a
4-planet Nice model simulation that meets these criteria but that
nevertheless fails to reproduce the current terrestrial system
due to a surprising resonant interaction with an ice giant. \\

\begin{figure}
\resizebox{\hsize}{!}{\includegraphics[angle=-90]{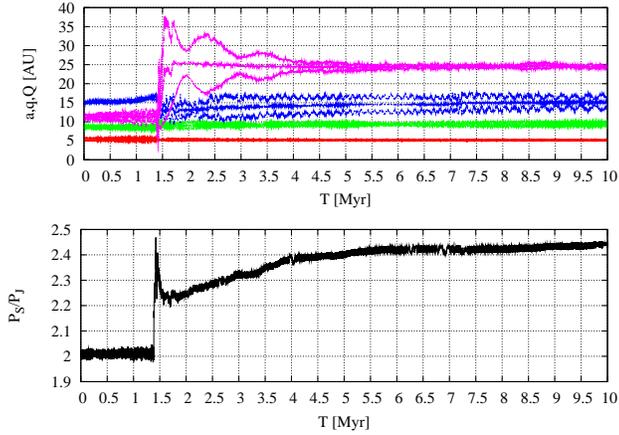}}
\caption{Evolution of the semi-major axis, perihelion and aphelion of
  the giant planets (top panel) and the Jupiter-Saturn period ratio
  (bottom panel) for the jumping Jupiter simulation used
    here. Red is Jupiter, green is Saturn, blue is Uranus and magenta
  is Neptune.}
\label{gps4pl}
\end{figure}

The { first 10~Myr of} the evolution of the giant planets in the test 4-planet Nice model simulation is displayed in
Fig.~\ref{gps4pl}. { The simulation was run for 100~Myr but after 10~Myr the gas giants had settled on their final orbits and the
migrating ice giants have little effect on the terrestrials.} At the end of the migration Uranus and Neptune are closer to Jupiter
and Saturn than they are today, which increases the precession frequencies of all the giant planets. The period ratio of the gas giants
jumps to about 2.23 and then increases to beyond 2.4 within 4~Myr, so that $\tau \sim$ 1-1.3~Myr. Thus, the simulation appears to be
compatible with the conditions we imposed from the smooth migration experiments: i) a jump
to approximately 2.25 or higher, and ii) subsequent smooth migration on a time scale of $\tau \sim$ 1~Myr. { This simulation
satisfies almost all conditions imposed by Nesvorn\'{y} (2011) and Nesvorn\'{y} \& Morbidelli (2012): only the final semi-major axis
of Uranus is too low and its inclination is too high. Their other criteria -- having four planets at the end, having the amplitude of
the Jovian eccentricity eigenmode $e_{55}>0.022$, having $P_S/P_J$ jump from $<$2.1 to $>$2.3 in a time span shorter than 1~Myr -- are
all matched. The final semi-major axes are 5.15, 9.32, 15.2 and 24.5, eccentricities are 0.027, 0.073, 0.042 and 0.009, inclinations
with respect to the invariable plane are 0.54$^\circ$, 1.78$^\circ$, 2.43$^\circ$ and 0.55$^\circ$, and $e_{55}=0.0342$, close to 80\%
of the current value. The amplitude of the latter bears direct correlation to the dynamical excitation of the terrestrial planets
(Brasser et al., 2009; Agnor \& Lin, 2012).}.\\

\begin{figure}
\resizebox{\hsize}{!}{\includegraphics[angle=-90]{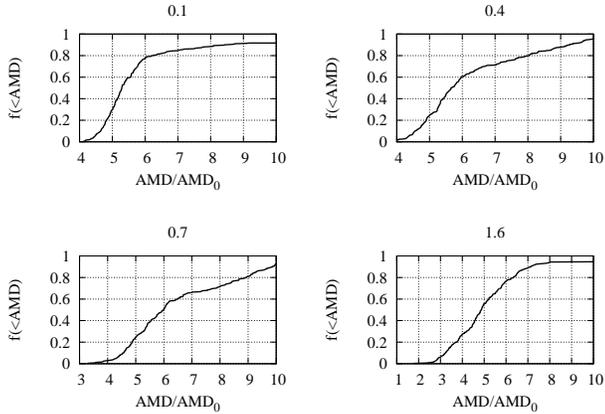}}
\caption{Cumulative distributions of the AMD of the terrestrial
  planets after the migration of the giant planets for various initial
  AMD values. The evolution of the giant planets is displayed in
  Fig.~\ref{gps4pl}.}
\label{amdtp4pl}
\end{figure}

In Fig.~\ref{amdtp4pl} we plot the cumulative distribution of the AMD
of the terrestrial planets after the migration of the giant planets
for various initial values of the AMD. The headers above each panel
depict the initial fraction of the current AMD. There are several
interesting features. First, even when the terrestrials are originally
very cold (AMD = 0.1), the minimum AMD after migration is more than 3
times the current value, with a median value near 4. Increasing the
initial AMD also increases the final median value. Second, the AMD
almost always increases, and thus the probability of destructive
interference -- which causes an overall reduction in the AMD -- is
much lower than the $\sim$10\% found by Brasser et al. (2009) and
Agnor \& Lin (2012). In order to understand why the AMD increases by
such a large amount, we plot an example of the evolution of the
terrestrials below.\\

\begin{figure}
\resizebox{\hsize}{!}{\includegraphics{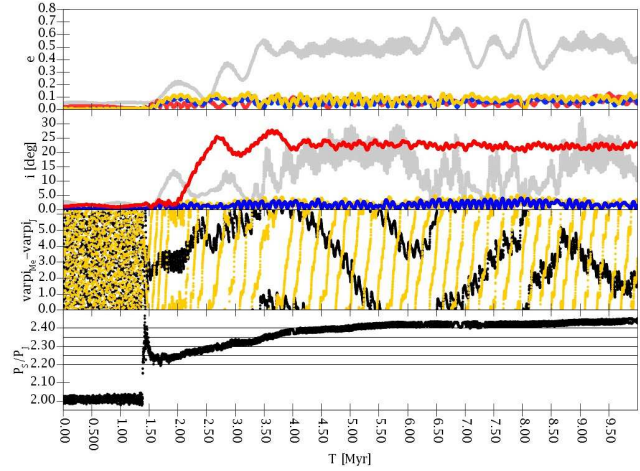}}
\caption{Evolution of the eccentricities (top), inclinations
  (second-from-top), $\varpi_{\rm{Me}}-\varpi_J$ (grey) and
  $\varpi_V-\varpi_J$ (yellow) (third-from-top) and $P_S/P_J$ (bottom
  panel). The colour coding is grey for Mercury, yellow for Venus,
  blue for Earth and red for Mars.}
\label{evo1}
\end{figure}

Figure~\ref{evo1} shows the evolution of the eccentricity and
inclination of the terrestrial planets (top and top-middle
panels). The colour coding is grey for Mercury, yellow for Venus, blue
for Earth and red for Mars. In the bottom-middle panel we plot the
angles $\Delta \varpi_{\rm{Me-J}}= \varpi_{\rm{Me}}-\varpi_{\rm{J}}$
in black and $\Delta \varpi_{\rm{V-J}} =
\varpi_{\rm{V}}-\varpi_{\rm{J}}$ in yellow. The bottom panel shows the
evolution of $P_S/P_J$. Immediately after the instability the argument
$\Delta \varpi_{\rm{Me-J}}$ librates with a long period. This secular
resonance between Jupiter and Mercury substantially increases
Mercury's eccentricity before it settles near 0.6. The increase in
Mercury's inclination from roughly 1$^\circ$ to approximately
10$^\circ$ is caused by a secular resonance with Venus: the argument
$\Omega_{\rm{Me}}-\Omega_{\rm{V}}$ changes the direction of
circulation. The later increase at $t=8$~Myr is also caused by
interaction with Venus.\\

The second feature is the sudden increase in Mars' inclination at
$t=1.5$~Myr. This increase is caused by a secular resonance with
Uranus (not shown): the argument $\Omega_{\rm{Ma}}-\Omega_{\rm{U}}$
slows down, librates around 0$^\circ$ for one oscillation with period
1~Myr and then continues to circulate. During this libration Mars
experiences its rapid increase in its inclination. The temporary
coupling between Uranus and Mars demonstrates that the evolution of
the ice giants could play an important role in shaping the secular
architecture of the inner planets: during the migration phase, when
the eccentricities and inclinations of the ice giants are much higher
than they are now, they may interact with the terrestrial planets
directly. It also demonstrates the importance of the ice giants
obtaining their current orbits at the end of the migration phase,
which this simulation does not adequately do.\\

{ One aspect that merits discussion is how our results depend on the amplitude of the Jovian eccentricity eigenmode $e_{55}$.
Jupiter's eccentricity forcing is present in the eccentricities of all the terrestrial planets (e.g. Brouwer \& van Woerkom, 1950).
The amplitude of Jupiter's forcing term on the terrestrials is directly proportional to $e_{55}$ itself. In this simulation the final
amplitude is lower than the current one, so that we would expect the final terrestrial AMD to be lower than its current value. Yet the
strange behaviour of Mercury and Mars during the migration substantially increases the final AMD of the terrestrial system. A lower
amplitude of the Jovian $e_{55}$ mode could have decreased the final AMD but it would be inconsistent with the current secular
architecture of the Solar System and not be sufficient to compensate for the increased AMD values of Mercury and Mars. In conclusion,
the evolution of the giant planets needs to excite $e_{55}$ to a value comparable to the current one without any of the terrestrial
planets getting caught in a secular resonance.}\\

{ As mentioned previously this particular simulation satisfies all of the constraints laid out in the models of Nesvorn\'{y}
(2011) and Nesvorn\'{y} \& Morbidelli (2012) for giant planet migration. Therefore it may be considered as a best case scenario for a
jumping Jupiter evolution of the giant planets in regards to their influence on the terrestrial planets.}\\

This simulation demonstrates that Mercury and Mars are more
susceptible than Venus and Earth to the evolution of the outer
planets. Of course the Earth (and Venus) also suffer the same secular
resonance with Uranus that Mars does because $s_3 \sim s_4$, but
because of its larger inertia the Earth's share of the AMD increases a
lower amount than Mars'. \\

From the above results it appears that the migration of the giant
planets should proceed on an even shorter time scale, with little to
no migration of Jupiter and Saturn occurring after the
jump { and keeping the $e_{55}$ mode at a value lower than or equal to the current one.}
Similarly, a jump only to 2.25 may have aggravated this 
particular case, and a jump to 2.3 may be essential -- as was
demonstrated in the previous section. Nesvorn\'{y}
\& Morbidelli (2012) show that this is very unlikely for a 4-planet
case (with probability lower than 1\%). However, we found one such
case with initially 5 planets in the simulations of Nesvorn\'{y} \&
Morbidelli (2012). Thus in the next subsection we present the outcome
of a 5-planet Nice model run that matches our criteria.

\subsection{A 5-planet Nice model simulation}
\begin{figure}
\resizebox{\hsize}{!}{\includegraphics[angle=-90]{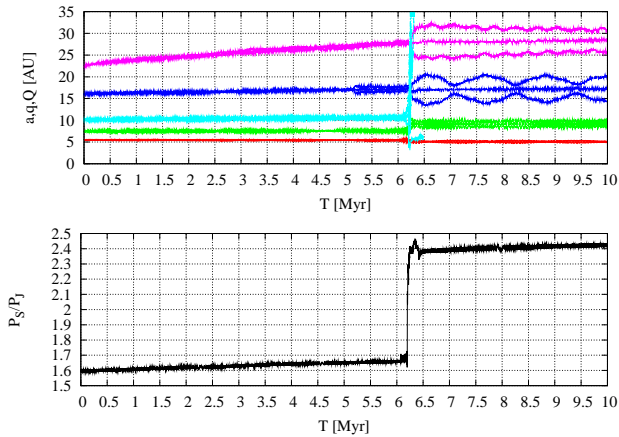}}
\caption{Evolution of the semi-major axes, perihelia and aphelia of a
  5-planet Nice model simulation (top panel) and the Jupiter-Saturn
  period ratio (bottom panel). Red is Jupiter, green is Saturn, blue
  is Uranus and magenta is Neptune. The colour for the 5th planet is
  cyan.}
\label{gps5pl}
\end{figure}

\begin{figure}
\resizebox{\hsize}{!}{\includegraphics[angle=-90]{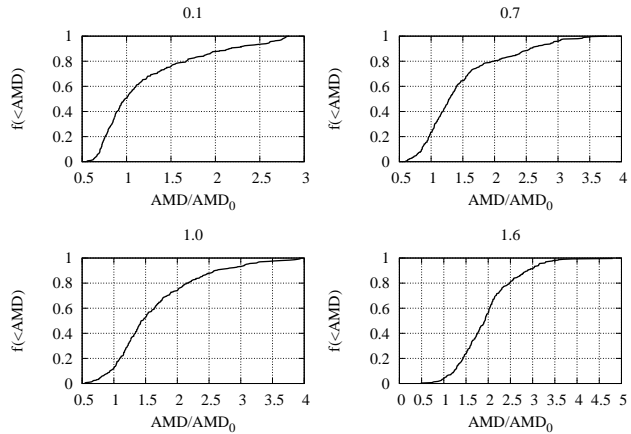}}
\caption{Same as Fig.~\ref{amdtp4pl} but now after the 5-planet Nice model simulation.}
\label{amdtp5pl}
\end{figure}
In this subsection we report on the results of exposing the
terrestrial system to a 5-planet Nice model simulation, which was
taken from Nesvorn\'{y} \& Morbidelli (2012). The planets started in a
quintuple resonant configuration: 3:2,3:2,2:1,3:2. The extra ice
giant's mass was equal to that of Neptune and the planetesimal disc
mass was 20~$M_{\oplus}$. The evolution of the system { for the first 10~Myr}
is displayed in Fig.~\ref{gps5pl}. The innermost ice giant (cyan) is ejected after
6.3~Myr, right after it is first scattered inwards by Saturn and then immediately
ejected by Jupiter, resulting in the big jump in $P_S/P_J$. There is
very little subsequent migration of the gas giants because the mass of
the planetesimal disc outside of Neptune was much lower than in the
4-planet case (Nesvorn\'{y} \& Morbidelli, 2012). { At this stage} both Uranus and
Neptune end up too close to the Sun, although they are farther than in
the 4-planet case. { This simulation satisfies almost all conditions imposed by Nesvorn\'{y} (2011) and Nesvorn\'{y} \& Morbidelli
(2012): only the final eccentricity of Uranus is too high. Their other criteria are all satisfied. The final semi-major axes are 5.06,
9.16, 17.2 and 28.2, eccentricities are 0.015, 0.061, 0.171 and 0.082, inclinations with respect to the invariable plane are
0.58$^\circ$, 1.50$^\circ$, 1.08$^\circ$ and 0.63$^\circ$, and $e_{55} \sim 0.04$, close to the current value. We want to
stress that we have only used the first 10~Myr of this simulation because later stages only featured slow migration and damping of
Uranus and Neptune which is unlikely to strongly affect the terrestrial planets. At the end of the simulation, all criteria from
Nesvorn\'{y} (2011) and Nesvorn\'{y} \& Morbidelli (2012) are satisfied.}\\

In our simulations we compute the inclinations of all planets with respect to the invariable plane { at the
beginning of the simulation. The mutual scattering of the giant planets and the ejection of an ice giant changes the total angular
momentum vector and thus the invariable plane.} We checked the simulations for a sudden jump in the inclinations of the terrestrial
planets when the first ice giant was ejected but witnessed no such behaviour. { In addition, dynamical friction from the
planetesimals in the original simulations damped the inclinations of the giant planets, which also changes the invariable plane. Thus,
for simplicity, we pinned the inclinations to the invariable plane at the beginning of the simulation.}\\

The AMD response of the terrestrial system to the evolution of the
giant planets is depicted in Fig.~\ref{amdtp5pl}. This figure should
be compared to Fig.~\ref{amdtp4pl} for the 4-planet case. One may see
that the AMD excitation in this simulation is much lower than for the
4-planet case. The top-left panel shows that for a low AMD (0.1) the
median final AMD equals the current value. Even when the initial AMD
was equal to the current value, it is still reproduced 10\% of the
time. Though the smooth migration simulations suggested that the
primordial AMD of the terrestrial planets had to be very low, the
outcome of the 5-planet case shows that the current AMD can be
reproduced with a reasonable probability if the primordial value was
as high as 70\% of the current one (reproduced $\sim$20\% of the
time). The results from this simulation also demonstrate that
destructive interference occurs at most with a $\sim$10\% probability
(Brasser et al., 2009; Agnor \& Lin, 2012), but is not strong enough
to reduce an initially higher AMD ($>$ 1.0) to the current
system (AMD $\sim$ 1.0) with a reasonable ($>$10\%) probability (see the
bottom right panel of Fig.~\ref{amdtp5pl} for an example of an initial
AMD of 1.6).\\

\begin{figure}
\resizebox{\hsize}{!}{\includegraphics[angle=-90]{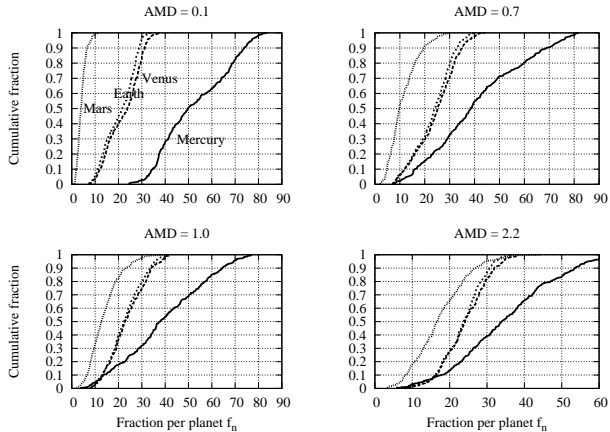}}
\caption{Cumulative distributions of the fraction of the total AMD in each terrestrial planet, $f_n$ -- see equation (\ref{fracp}).
The headers above the panels indicate the amount of original AMD.}
\label{amdfrac5pl}
\end{figure}
How is the AMD shared among the terrestrial planets after the migration of the giant planets: does the share in each planet stay
roughly the same, or are they substantially altered? { After each simulation we record the average final AMD and the average share
or fraction of the total in each planet, $f_n$ -- see equation (\ref{fracp}). We plot the cumulative distribution of $f_n$ for
various initial AMD values in Fig.~\ref{amdfrac5pl}. This figure serves to illustrate the spread of the share in each planet and
thus the amount of excitation relative to the other planets.} The value of the primordial AMD are shown above the panel. The fraction
of the AMD in Venus or Earth is very similar due to their strong coupling: { for all simulations the share of the AMD in both Venus and
Earth ranges from 10\% to about 40\%. { The share in Mars increases with increasing AMD, and it typically has between 0 to
20\%. The spread in Mercury is much larger than that of the other planets, though it does decrease slightly with increasing AMD. In
half of all simulations 50\% of the AMD of the terrestrial system is taken up by Mercury, so that the other three planets combined also
have 50\% and are thus dynamically colder. Most likely at higher AMD some of the excess from Mercury is transferred to Mars. However,
for all intents and purposes the final share in each planet appears independent of the original AMD value. } For the cold initial
system (AMD=0.1) Mars' value is systematically much lower than found today and Mercury's is mostly higher. This suggests that Mars
experiences little change in its orbit and loses AMD to the other planets because the other three planets gain AMD. Mercury,
especially, must have been forced by some mechanism. { For higher initial AMD all planets can exchange AMD with each other and the
relative forcing of the inner three during the instability compared to Mars is smaller, hence Mars' share appears to increase with
increasing AMD.}\\

When examining the evolution of the giant planets and comparing it
with the terrestrial planets we find that Mercury gets caught in the
resonance $g_7=g_1$: after the innermost ice giant is ejected
$g_7 \sim$ 5.3~$\arcsec$/yr, which is close to $g_1$
(5.5~$\arcsec$/yr). For reference, currently $g_7 \sim
3.1$~$\arcsec$/yr. To examine the severity of this secular resonance
we plot in Fig.~\ref{eve5pl} the averaged minimum, mean and maximum
eccentricities of Mercury (top-left) and Venus (top-right) as a
function of the initial AMD. Also plotted is the probability of
their maximum eccentricities remaining below 0.35 and 0.09
respectively (bottom-left for Mercury and bottom-right for Venus). For
low to mid-AMD values Venus remains low but for all AMD values
Mercury's mean eccentricity is $\sim$0.3 and the probability of it
being below 0.35 is just over 50\%. This is acceptable given that
Uranus' final position is too close to the Sun. Subsequent chaotic
diffusion may lower Mercury's eccentricity to its current value
(Laskar, 2008).\\

{ Was there any way to avoid Mercury's high eccentricity? We have stated earlier that we have only used the first 10~Myr of the
simulation. Having run for longer would not have solved the issue of Mercury's high eccentricity because the secular resonance crossing
with Uranus would still have occurred on a similar time scale ($\sim$10~Myr) and yielded a similar increase in Mercury's eccentricity.}
\\

It appears that the giant planet evolution presented in Fig.~\ref{gps5pl} is mostly capable of reproducing the current AMD of the
terrestrial planets with a reasonable probability ($\sim$20\%), provided the initial AMD remained below 70\% of the current value. If
the AMD had been equal to the current value the probability for it to remain unchanged is approximately 10\%. The lowest initial AMD
value reproduced the current AMD of the terrestrial planets in 50\% of the simulations. { The simulation has more difficulty in
reproducing the current fractions of AMD in each planet. However,} Mercury's high eccentricity in these simulations is an artefact of
Uranus ending up too close to the Sun. { Mars is more difficult. The inner three planets are forced more during the instability
than Mars itself, so that the gain of AMD of the inner three occurs at the expense of Mars. Its final low share of the system AMD could
be increased to its current value if its original share had been higher than the other three planets i.e. if we had used a different
sharing among the planets at the beginning of our simulations. The question then becomes how high this initial share can be pushed
and whether we consider a final share of $<$10\% to be a successful outcome.} In the next section, we compare the outcome of our
simulations with those of terrestrial planet formation simulations and discuss its implications.

\begin{figure}
\resizebox{\hsize}{!}{\includegraphics[angle=-90]{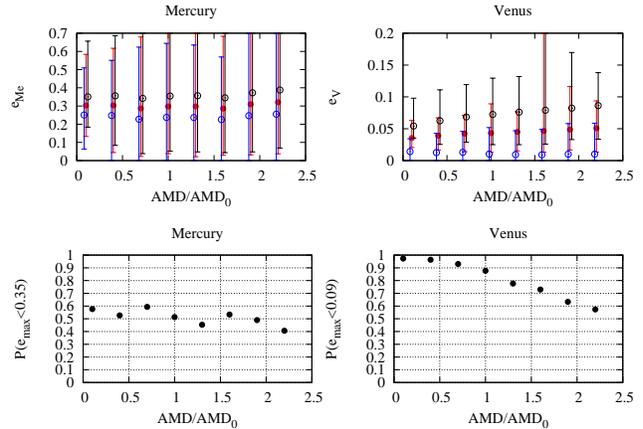}}
\caption{Averaged minimum, mean and maximum eccentricity for Mercury
  (top-right) and Venus (top-left) as a function of initial AMD.  The
  bottom panels depict the probability of Mercury's eccentricity
  staying below 0.35 (left) and Venus' below 0.09 (right).}
\label{eve5pl}
\end{figure}

\section{Comparison with terrestrial planet formation simulations and implications}
{ We have set up a high number of fictitious terrestrial systems and exposed them to the evolution of the migrating giant planets.
We have assumed that this migration of the giant planets occurred late and the terrestrial planets had already formed.} Some of our
results require further explanation, which we do here. We also compare our results with the outcome of
terrestrial planet formation simulations.\\
\begin{figure}
\resizebox{\hsize}{!}{\includegraphics[angle=-90]{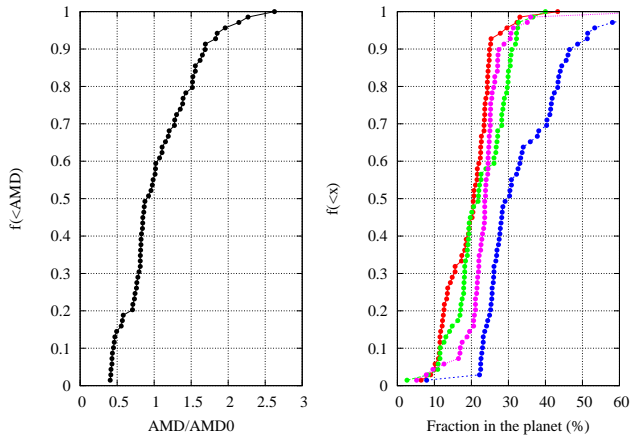}}
\caption{Cumulative value of the final AMD of the 4-planet terrestrial
  planets systems of Walsh et al. (2011) (right panel). The curve is
  taken from combining the results of several simulations. The left
  panel shows the cumulative distribution of the share of the AMD in
  each planet. Red bullets are the innermost planet, blue and green
  are for the middle two planets and the magenta corresponds to the
  outermost planet.}
\label{hwamd}
\end{figure}
First, { assuming late migration} we suggest that the primordial AMD of the terrestrial planets
was not only lower than its current value but also that most, or all,
of it was contained in Mars. Agnor \& Lin (2012), while not
investigating the change in AMD directly, also concluded that the
primordial orbit of Mars (and Mercury) had to be more excited than
those of Venus and Earth. Is this outcome, and the lower AMD that we
advocate here, consistent with terrestrial planet formation
simulations? To date the simulations best reproducing the
current mass--semi-major axis distribution of the terrestrial planets
are those of Hansen (2009) and Walsh et al. (2011). Hansen
(2009) ran numerical simulations of the formation of the terrestrial
planets by placing 400 equal-mass embryos in an annulus from 0.7~AU to
1~AU. The total mass of the embryos was 2~$M_{\oplus}$. Walsh et
al. (2011) used the migration of Jupiter and Saturn to reproduce
the outer edge of the planetesimal disc and recreate the initial
conditions of Hansen (2009). Unlike Hansen (2009) Walsh et
al. (2011) included planetesimals in their simulations.\\

We took the data from Walsh et al. (2011) and only kept the
cases with four terrestrial planets. We simulated these systems for
1~Myr with SWIFT MVS (Levison \& Duncan, 1994) to obtain their
averaged AMD, the share of the AMD in each planet and the partitioning
of the AMD among eccentricity and inclination. The results are
displayed in Fig.~\ref{hwamd}. The cumulative AMD, normalised to the
current value, is displayed in the left panel, where the plot
was generated by combining the results of several simulations. In the
right panel the share of the AMD in each planet is depicted. The
red dots correspond to the innermost planet, followed by green and
blue while the magenta dots pertain to the outermost planet. It
appears that the share of the AMD in each planet typically ranges from
10\% to 40\%, with a median around 25\%. This strongly suggests that
during the formation process the system reaches angular momentum
equipartition. For the current solar system these fractions are
34\% in Mercury, 20\% in Venus, 21\% in Earth and 25\% in Mars. Thus
the current partitioning among the terrestrial planets in the current solar 
system is consistent with their primordial ones, though it could be 
argued that Mercury's share reflects some additional excitation.\\

However, the primordial AMD sharing that would most likely lead to the 
current system (a higher share in Mars than in the other planets), does not 
closely match the results of Walsh et al. (2011) (Fig.~\ref{hwamd}). The lowest
system AMD from the Walsh et al. (2011) simulations (Fig.~\ref{hwamd} left panel) 
is approximately 40\% of the current one. Such a low primordial AMD has been shown in this work to have 
a high probability of not exceeding the current value following { late} giant planet migration.\\

The higher share of Mars is of more concern. { In the previous section we demonstrated that cases with a very low AMD yield a share
in Mars that is inconsistent with the results presented in Fig.~\ref{hwamd}, and thus we are inclined to reject a very low primordial
AMD of the terrestrial planets.} However, Walsh et al. (2011) included planetesimals in their simulations, and these damp the
eccentricities of the terrestrial planets through dynamical friction (e.g. O'Brien et al., 2006). Therefore the formation of a
lower-AMD system with a nearly-circular and nearly-coplanar Mercury, Venus and Earth and slightly eccentric and inclined Mars cannot be
ruled out if the original mass in planetesimals was higher than that in embryos and more confined to a narrower region inside of Mars.
It is likely that Mars is a stranded planetary embryo (e.g. Dauphas \& Pourmand, 2011) and both Mercury and Mars probably ended up at
their current positions by encounters with Venus and Earth (Hansen, 2009; Walsh et al., 2011). These encounters increased their
eccentricities and inclinations, and their subsequent isolation from other planetary embryos explains both their small size and
continued excited orbits. The lower density of small bodies at these planets' orbits would have decreased the amount of dynamical
friction they experienced and thus their orbits remained hotter than those of Venus and Earth { and their AMD share could have also
remained higher}. New terrestrial planet formation simulations need to explore whether such an outcome is possible.\\

Second, one could ask the question whether we could have done anything
differently. Are the Nice model simulations that we used
representative of what happened at that time? Nesvorn\'{y} (2011) and
Nesvorn\'{y} \& Morbidelli (2012) developed a set of criteria of what
they deemed a successful Nice model simulation. As stated earlier, they used both four
and five giant planets. They reported that to fulfil all of their criteria with a
four-planet case the probability is lower than 1\%, while
in the five-planet case the probability is at most 5\%. Thus, in
choosing our simulations we decided to opt for the one that gave the
best possible jump of Jupiter, while relaxing the criterion { of all planets ending at their current semi-major axes.
Our choice of simulations are a best case in which the AMD excitation of the terrestrials may be minimised. We justify this choice by
repeating that our goal was to try to determine whether the current excitation could be reproduced by late planet migration. To that
end the 5-planet case appears to have succeeded.} \\

At minimum the period ratio $P_S/P_J$ needs to jump from
$<2.1$ to $>2.3$ but still stay below 2.5. { This constraint holds both for early (Walsh \& Morbidelli, 2011) and late migration
(Brasser et al., 2009; Agnor \& Lin, 2012).} It is likely
that the primordial period ratio was $P_S/P_J \sim 1.5$ (Morbidelli et
al., 2007; Pierens \& Nelson, 2008), and thus the jump may have needed
to proceed from a period ratio of $\sim$ 1.5 to $>2.3$. This jump
occurred in the 5-planet simulation presented here but it is
substantial and difficult to accomplish by encounters with an ice
giant, even when the latter is ejected (Nesvorn\'{y} \& Morbidelli,
2012).\\

Unfortunately, { during late migration} the terrestrial planets are not entirely
unaffected by the ice giants' evolution. Despite the ice giants'
influence we argue that the final outcome may not change substantially
when the ice giants undergo a different evolution: the largest
threat to the stability of the terrestrial planets is the sweeping of
$g_5$ through the terrestrial region. Therefore, any simulation keeping
(all) four planets may be satisfactory, provided that the period ratio
$P_S/P_J$ undergoes the required jump to 2.3 in a short enough
time scale ($\tau \sim 0.1$~Myr) and suffers little migration
afterwards (Nesvorn\'{y} \& Morbidelli, 2012).\\

Another issue that warrants a discussion are the initial conditions of
the terrestrial planets. For the sake of simplicity we have taken the 
system AMD to be the independent variable rather than the
eccentricity and inclination of each planet. We randomised the phases
and ran a large number of simulations to determine a range of
outcomes. We performed a statistical analysis rather than a
case-by-case investigation. However, we also decided to partition the 
$z$-component of the angular momentum $h_z
= \sqrt{1-e^2}\cos i$ evenly among inclination and eccentricity,
inspired by the results of the simulations by Hansen (2009) and Walsh
et al. (2011). Would the results have changed substantially if we had
done this differently e.g. by setting $e=\frac{3}{2}\sin i$? The AMD
partitioning is altered by the migration of the giant planets and thus
the final sharing of the AMD can be changed by applying a different
primordial distribution among the planets. Thus we think that our
choice of setting $e = \sin i$ is justified. \\

The last issue pertains to the sharing of the AMD among the
terrestrial planets. In our simulations we decided to keep each
planet's current share but we could have made this a random variable
as well, limiting it to the range displayed in the right panel of
Figs.~\ref{hwamd}. Once again we opted for simplicity in using the
current values. For low initial AMD the shares of Mercury, Venus and
Earth show a reasonable dispersion at the end of the 5-planet
simulation. The only planet whose share remains low is Mars, but we
can use the argument above that it was originally hotter than the
other planets { to offset its final low AMD.} The conclusion of Mars being originally dynamically hotter was also reached by Agnor
\& Lin (2012). Changing the original sharing will add an extra layer of
complexity to the problem that becomes somewhat speculative and it may
no longer be possible to make any predictions about the original
AMD. Thus we decided to use the current partitioning. { However, Mars' low final share does suggest that very low primordial
AMD values may not be consistent with excitation of the AMD by late giant planet migration.}

\section{Conclusion}
In this study we investigated in detail how the orbits of the
terrestrial planets change when the giant planets undergo their late
instability. Based on criteria developed in Brasser et al. (2009),
Agnor \& Lin (2012) and Nesvorn\'{y} \& Morbidelli (2012) we subjected
the terrestrial planets to two jumping Jupiter Nice model
simulations. Thus the migration of Jupiter and Saturn occurs on a time
scale $\tau < 0.1$~Myr { and were best-case scenaria for both a 4-planet and 5-planet evolution, in which the duration of the
jump of Jupiter was the shortest.} 
We chose the initial AMD of the terrestrial planets as the independent variable and ran many simulations
with random phasing of the orbital angles of the terrestrial
planets. We recorded the final AMD of the terrestrial system and the
sharing of the AMD among the terrestrial planets.\\

From the numerical simulations that we performed three basic
conclusions are drawn. First, if the { late} giant planet migration
scenario that we subjected the terrestrial planets to are
representative of the reality, and to reproduce the current AMD with a
reasonable probability ($\sim$20\%), the primordial AMD of the
terrestrial planets should have been lower than 70\% of the current
value. If the primordial AMD had been higher than the current value,
the probability of the terrestrial planets ending up with their
current AMD value becomes very low ($<$1\%). { However, a very low primordial AMD ($\sim$0.1) can probably be ruled out because the
final share in Mars is inconsistent with the results of Walsh et al. (2011).}\\ 

Second, at present the terrestrial planets carry approximately equal
amounts of the system's AMD. { Assuming that their orbits were influenced 
by late giant planet migration} the primordial partitioning among the planets must
have been peculiar because the evolution of the giant planets left
Mars' orbit mostly intact; hence most (or perhaps all) of the
primordial AMD was taken up by Mars { because after migration it lost most of its share to the other planets}. 
From this configuration and the initially low AMD (0.4-0.7) we predict that Mars' primordial eccentricity and
inclination were similar to their current values. Mercury, Venus and Earth 
were approximately circular and coplanar. The low primordial AMD and the 
peculiar partitioning impose a new constraint for terrestrial planet formation
simulations.\\

Third, the Jupiter-Saturn period ratio had to have jumped from
$\sim$1.5 to beyond 2.3 with little subsequent migration. Having
little subsequent migration is needed to keep the final terrestrial
planet AMD at the current value when starting from a lower one. This
evolution of the giant planets is better realised with 5 planets than
with 4 (Nesvorn\'{y} \& Morbidelli, 2012).

\section{Acknowledgements}
\footnotesize{The Condor Software Program (HTCondor) was developed by the Condor Team at the Computer Sciences Department of the
University of Wisconsin-Madison. All rights, title, and interest in HTCondor are owned by the Condor Team. Support for KJW came
from the Center for Lunar Origin and Evolution of NASA's Lunar Science Institute at the Southwest Research Institute in Boulder, CO,
USA. We thank an anonymous reviewer for valuable feedback that improved this manuscript.}\\

\section{References}
Agnor C.~B., Lin D.~N.~C., 2012, ApJ, 745, 143\\
Batygin K., Brown M.~E., Betts H., 2012, ApJ, 744, L3\\
Bottke W.~F., Vokrouhlick{\'y} D., Minton D., Nesvorn{\'y} D., Morbidelli 
A., Brasser R., Simonson B., Levison H.~F., 2012, Natur, 485, 78 \\
Brouwer, D., van Woerkom, A. J. J. 1950. Astron. Papers Amer. Ephem. 13, 81-107.\\
Brasser R., Morbidelli A., Gomes R., Tsiganis K., Levison H.~F., 2009, A\&A, 507, 1053\\
Correia A.~C.~M., Laskar J., 2010, Icar, 205, 338\\
Dauphas N., Pourmand A., 2011, Nature, 473, 489\\
Fernandez, J.A., Ip, W.-H., 1984, Icarus 58 109\\
Gomes R., Levison H.~F., Tsiganis K., Morbidelli A., 2005, Natur, 435, 466 \\
Hahn, J. M., Malhotra, R., 1999, The Astronomical Journal 117, 3041\\
Hansen B.~M.~S., 2009, ApJ, 703, 1131\\
Laskar J., 1997, A\&A, 317, L75 \\
Laskar J., 2008, Icar, 196, 1\\
Levison H.~F., Duncan M.~J., 1994, Icar, 108, 18\\
Minton, D. A., Malhotra, R., 2009, Nature 457, 1109\\
Morbidelli A., Levison H.~F., Tsiganis K., Gomes R., 2005, Natur, 435, 462\\
Morbidelli A., Tsiganis K., Crida A., Levison H.~F., Gomes R., 2007, AJ, 134, 1790\\
Morbidelli A., Brasser R., Tsiganis K., Gomes R., Levison H.~F., 2009, A\&A, 507, 1041\\ 
Morbidelli A., Brasser R., Gomes R., Levison H.~F., Tsiganis K., 2010, AJ, 140, 1391\\
Nesvorn{\'y} D., Vokrouhlick{\'y} D., Morbidelli A., 2007, AJ, 133, 1962\\
Nesvorn{\'y} D., 2011, ApJ, 742, L22\\
Nesvorn{\'y} D., Morbidelli A., 2012, AJ, 144, 117\\
Nesvorn{\'y} D., Vokrouhlick{\'y} D., Morbidelli A., 2013, ApJ, 768, 45\\
Nobili A., Roxburgh I.~W., 1986, IAUS, 114, 105\\
O'Brien D.~P., Morbidelli A., Levison H.~F., 2006, Icar, 184, 39\\
Raymond S.~N., O'Brien D.~P., Morbidelli A., Kaib N.~A., 2009, Icar, 203, 644\\
Pierens A., Nelson R.~P., 2008, A\&A, 482, 333\\
Petit J.-M., Morbidelli A., Chambers J., 2001, Icar, 153, 338\\
Thommes, E.W., Duncan, M. J., Levison, H. F., 1999, Nature 402, 635\\
Tsiganis, K.; Gomes, R.; Morbidelli, A.; Levison, H. F., 2005, Nature 435, 459\\
Saha P., Tremaine S., 1994, AJ, 108, 1962\\
Walsh K.~J., Morbidelli A., Raymond S.~N., O'Brien D.~P., Mandell, A.~M., 2011, Nature, 475, 206\\
Ward W.~R., 1981, Icar, 47, 234
\end{document}